\documentclass{ecai}  % use option [doubleblind] for double blind submission and hiding the authors section

\usepackage{graphicx}
\usepackage{latexsym}
\usepackage{xcolor}
\usepackage{amsfonts}
\usepackage{booktabs} % for professionexa AI tables
\usepackage{graphicx}
\usepackage{lipsum}
\newcommand\scalemath[2]{\scalebox{#1}{\mbox{\ensuremath{\displaystyle #2}}}}

\paperid{612}        % paper id for double blind submission

\makeatletter
\def\blfootnote{\xdef\@thefnmark{}\@footnotetext}
\makeatother

\begin{document}

\begin{frontmatter}

  \title{SCRAPS: Speech Contrastive Representations of Acoustic and Phonetic Spaces}

\author[1]{ \fnms{Ivan}~\snm{Valles-Perez}\thanks{Corresponding Author Email: ivallesp@amazon.com.}}
\author[1]{ \fnms{Grzegorz}~\snm{Beringer}}
\author[1;2] {\fnms{Piotr}~\snm{Bilinski}}
\author[1]{ \fnms{Gary}~\snm{Cook}}
\author[1]{ \fnms{Roberto}~\snm{Barra-Chicote}}

\address[1]{ Alexa AI, Amazon}
\address[2]{ University of Warsaw}

\begin{abstract}
Numerous examples in the literature proved that deep learning models have the ability to work well with multimodal data. Recently, \textit{CLIP} has enabled deep learning systems to learn shared latent spaces between images and text descriptions, with outstanding zero- or few-shot results in downstream tasks. In this paper we explore the same idea proposed by \textit{CLIP} but applied to the speech domain, where the phonetic and acoustic spaces usually coexist. We train a \textit{CLIP}-based model with the aim to learn shared representations of phonetic and acoustic spaces. The results show that the proposed model is sensible to phonetic changes, with a 91\% of score drops when replacing 20\% of the phonemes at random, while providing substantial robustness against different kinds of noise, with a 10\% performance drop when mixing the audio with 75\% of \textit{Gaussian} noise. We also provide empirical evidence showing that the resulting embeddings are useful for a variety of downstream applications, such as intelligibility evaluation and the ability to leverage rich pre-trained phonetic embeddings in speech generation task. Finally, we discuss potential applications with interesting implications for the speech generation and recognition fields.

\end{abstract}

\end{frontmatter}

\section{Introduction}
\label{submission}

\blfootnote{Work conducted when all authors were at Alexa AI, Amazon.}

Recent advances in deep learning have catapulted computer vision, improving the domain across multiple axes \cite{chai2021}. One of the most important breakthroughs is \textit{CLIP} (Contrastive Language-Image Pre-training) \cite{radford2021}, a deep learning model trained on vast amounts of multimodal data. In this case, by means of contrastive learning \cite{lekhac2020}, \textit{CLIP} showed the ability to learn to match images with written descriptions, and vice-versa. The representation power of this model at zero-shot regimes is such that it is able to surpass supervised models specifically trained for a determined task. The embeddings generated with \textit{CLIP} have also been proved useful for other downstream tasks involving image and/or descriptions, in form of pretrained embeddings with rich latent representations \cite{conde2021, rombach2021, khandelwal2022, urli2022}. In these settings, reusing \textit{CLIP} embeddings becomes a cornerstone for problems with small datasets. The extraordinary success of \textit{CLIP} and the simplicity of its approach motivates exploring other applications to other fields where multimodal data is essential. 

The field of speech technologies has come a long way in recent years, with deep learning models achieving impressive performance on a wide range of tasks \cite{vanderoord2016, shen2018, garcia2019}. However, there are still many challenges and opportunities in this domain \cite{georgila2017}, particularly when it comes to exploiting large amounts of data. On the speech generation side, one of the main difficulties is to build a model that correctly aligns the phonetic and acoustic sequences, leading to a natural prosody with fluent speech and high intelligibility, while still capturing the prosody variations \cite{ren2022}. On the opposite side, automatic speech recognition systems struggle with long-tail words recognition \cite{indra2020}, and speech vs background disentanglement \cite{kinoshita2020}. 

This work explores the use of \textit{CLIP} models for learning shared phonetic-acoustic latent spaces that can be later exploited for further downstream tasks.  The methodology described in this study, which we name \textit{SCRAPS} (Speech Contrastive Representation of Acoustic and Phonetic Spaces) offers a promising approach for learning such common representations. By learning phonetic-acoustic representations, we aim to enable the use of rich latent spaces in a variety of downstream tasks, such as cheap intelligibility evaluation, rich pre-trained embeddings, a differential intelligibility loss and on-the-fly training data filtering. This last example is specially relevant at the moment given the vast amounts of found data available online. 

We believe that this approach has the potential to help advancing the state of the art in speech generation and recognition towards learning from large-scale data, looking at the impact that \textit{CLIP} model has had in the computer vision domain. In the following sections, we will provide a detailed description of our proposed methodology, as well as experiments and results that backup our hypotheses, demonstrating its effectiveness. We also discuss potential downstream applications and use cases where the \textit{SCRAPS} model may be useful, together with some initial experiments, and suggest future research lines.

\section{Related work}
There are various examples in the literature with notable similarity to the work presented in this paper, one of the most relevant being \textit{CLAP} \cite{elizalde2022}. \textit{CLAP} stands for Contrastive Language-Audio Pretraining, and it is a \textit{CLIP} based model that is trained on text description-audio pairs with the aim of producing a shared multimodal representation of acoustic and semantic domains. Although arguably similar to this work, \textit{CLAP} uses the text from a semantic perspective, as the dataset used to train this model contents pairs of audio samples with corresponding textual descriptions. \textit{SCRAPS} is trained over pairs of audio clips containing speech utterances and their corresponding transcribed phonetic sequences. The objective of \textit{SCRAPS} is to learn a shared latent space between the phonetics and its acoustic realizations, hence the nature of the problem is orthogonal to \textit{CLAP}.

Another example of \textit{CLIP}-based approach that operates in the speech domain is \textit{SpeechCLIP} \cite{shih2022}. In that work, the authors build a system that learns shared representations of images with spoken captions. By means of \textit{CLIP}, the authors establish a third alignment with text descriptions. In \textit{M-SpeechCLIP} \cite{berry2022}, the authors study the usefulness of English pre-trained \textit{SpeechCLIP} to other languages and also propose multilingual versions of it. \textit{Wav2CLIP} \cite{wu2021} is another work with the same idea as \textit{SpeechCLIP}, with the difference that the authors use the \textit{CLIP} image encoder to distill an audio encoder that produces latent representations compatible with the images. \textit{AudioCLIP} \cite{guzhov2021} is another example where the original \textit{CLIP} model gets extended to the audio domain. The authors first pre-train the audio encoder using the frozen pre-trained \textit{CLIP} encoders, and later unfreeze all the modules to fine-tune the whole system together.  

These approaches differ from ours in the following points:
\begin{enumerate}
    \item The approaches described above operate with text descriptions and at word level to infer semantics (as they rely on \textit{CLIP} text encoder), while \textit{SCRAPS} operates in the phonemes space, which brings the problem closer to the speech generation/recognition field, and makes it more practical for those use-cases.  
    \item The models found in the literature (aside from CLAP) rely on the pre-trained \textit{CLIP} models (\textit{SpeechCLIP} at inference time and \textit{Wav2CLIP} and \textit{Audio2CLIP} at training time), while our solution is an end-to-end model, and all the models are directly optimized to work with phonetic and acoustic spaces. 
    \item \textit{SpeechCLIP} and \textit{Wav2CLIP} do not directly optimize the speech-text shared space, but use images as a common intermediate space. In practice, this means that the error inherent in the original \textit{CLIP} model (which operates on the image-text multimodality) becomes effectively irreducible, and it is added to the image-audio model error. 
\end{enumerate}

To the best of our knowledge, our work is the first attempt to use a \textit{CLIP}-based strategy to learn joint phonetic and acoustic spaces. For this reason, we believe this work is of high interest for the speech generation/recognition community.

\begin{figure}[h!]
	\centering
	\includegraphics[width=1.0\linewidth]{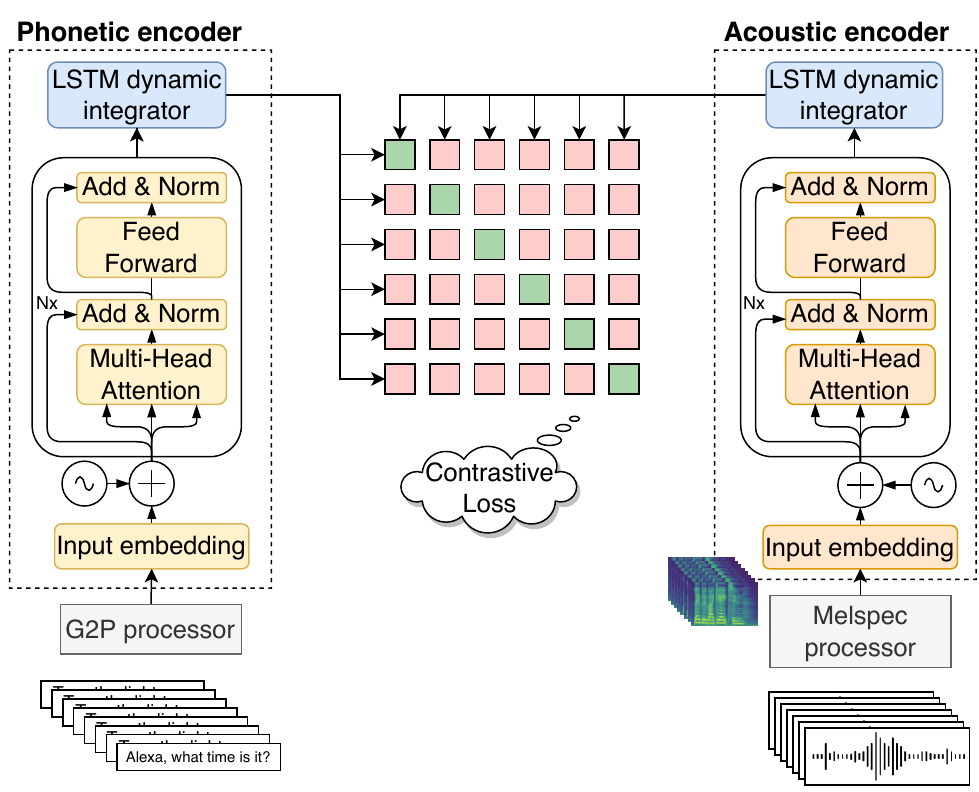}
	\caption{\textit{SCRAPS} model architecture, with a phonetic encoder in the left hand side, and an acoustic encoder in the right hand side.}
	\label{fig:scraps}
\end{figure}

\section{Methods}

\subsection{CLIP}
\textit{CLIP} is composed of two encoders, one for the images and one for the text. Each one receives the corresponding input (\textit{i.e.} the image is fed to the visual encoder and its description to the text encoder) and generates a real valued vector, one derived from the image $\vec{u} \in \mathbb{R}^D$, another derived from the text $\vec{v} \in \mathbb{R}^D$, both with the same dimensionality $D$. The objective of the network is to make both vector spaces compatible, such that corresponding image-text pairs produce, ideally, the same vectors – e.g. $(\vec{u}_1, \vec{v}_1)$ – and non-corresponding pairs produce dissimilar vectors $(\vec{u}_1 , \vec{v}_2 )$. This is achieved by maximizing the dot product between corresponding pairs of vectors, which matrix is represented as $\mathbf{L}$, with shape $N \times M$, \textit{i.e.} $N$ images times $M$ descriptions, and minimizing it for non-corresponding pairs. In the original form, $M=N$ as the positive and negative pairs are constructed using all the pairwise combinations of the elements in the \textit{minibatch}.

This is done by using a symmetric cross-entropy loss (cross-entropy across text and across images), as described in Equation \ref{eq:clip_loss}, where $\mathbf{L}$ represents the logits matrix resulting from the dot product of the outputs of the two encoders, and $\mathbf{P_1}$ and $\mathbf{P_2}$ are the two \textit{softmax}-normalized matrices across the two possible axes ($T$ represents the \textit{softmax} temperature). The pairing of images and descriptions is done using the elements of the \textit{minibatch}. 

\begin{equation}
	 \scalemath{0.9}{\mathbf{P_1}(i, j) = \frac{e^{\mathbf{L}(i, j)/T}}{\sum_{k=1}^N {e^{\mathbf{L}(k, j)/T}}},\ \  \mathbf{P_2}(i, j) = \frac{e^{\mathbf{L}(i, j)/T}}{\sum_{k=1}^M {e^{\mathbf{L}(i, k)/T}}}}.
\end{equation}

\begin{equation} \label{eq:clip_loss}
	\scalemath{0.9}{\mathcal{L} = \sum_{j=1}^M\sum_{k=1}^N \mathbf{I}(k,j)\log \mathbf{P_1}(k,j) + \sum_{i=1}^N\sum_{k=1}^M \mathbf{I}(i,k)\log \mathbf{P_2}(i,k)}.
\end{equation}

\subsection{SCRAPS}

\textit{SCRAPS} is an adaption of the original \textit{CLIP} idea to the speech domain. Consequently, the description encoder included in the original \textit{CLIP} architecture is replaced by a phonetic encoder, which receives a sequence of phonemes as input, and produces a vector as output. The image encoder is replaced by an acoustic encoder, which receives a mel-spectrogram as input and produces a vector as output. 

The architecture used for this study is represented in Figure \ref{fig:scraps}. As it can be seen in the figure, the model is built of two blocks: a phonetic encoder (left hand side) and an acoustic decoder (right hand side). Both modules take their corresponding input and produce a $D$ dimensional vector (time independent): $\mathbb{R}^D$. As shown in the diagram, the blocks contain two modules: a transformer backbone (inspired on transformer TTS \cite{naihan2019, vaswani2017}), and an \textit{LSTM} \cite{schmidhuber1997} on top. The role of the \textit{LSTM} is to integrate the representations coming from the transformers whose inputs have dynamic lengths. It produces a forward-rolling integration of the outputs vectors from the transformers. The last state of the \textit{LSTM} is used as the final \textit{SCRAPS} vector. The \textit{LSTMs} of the two encoders share weights to propagate the latent spaces compatibility back to the output of the transformer outputs (see section \ref{sec:emb} as an empirical proof of this assumption). That helps the time-dependent vectors live in the same space for both encoders, while still maintaining a time-dependent latent representation which may be used for future downstream tasks where the either of the \textit{SCRAPS} encoders is used as a pre-trained module. 

As it is shown in the middle of Figure \ref{fig:scraps}, the \textit{SCRAPS} model is trained with a contrastive loss where the objective is to maximize the scores of the matching pairs, while minimizing the scores of the non-matching pairs. This is achieved by computing the dot product of $\mathbf T \in \mathbb{R}^{B\times D}$, i.e. \textit{SCRAPS} vectors of the $B$ phonetic sequences of the \textit{minibatch}, and $\mathbf A\in\mathbb{R}^{B\times D}$, i.e. \textit{SCRAPS} vectors of the $B$ mel-spectrograms in the \textit{minibatch}: $\mathbf L=\mathbf T\mathbf A^T$;$\mathbf L\in \mathbb{R}^{N\times N}$. $\mathbf L$ is referred as the \textit{SCRAPS} scores matrix, hereafter. \textit{SCRAPS} is trained with the same loss as in the original \textit{CLIP} model (Equation \ref{eq:clip_loss}).
%The same loss used in the original \textit{CLIP} model (defined in Equation \ref{eq:clip_loss}) has been used during \textit{SCRAPS} training. 

\section{Experiments}
\subsection{Setup}
We have trained a \textit{SCRAPS} model on a large proprietary dataset containing 60,000 hours of de-identified speech data in US English. The data is composed of recordings of untrained speakers with variety of background noises, unnormalized pauses, and in some cases, even some samples with concurrent speech where one of the speakers dominates over the others. Each recording is accompanied with its corresponding transcription. The phonetic sequences are derived from the text transcriptions by using a simplified grapheme to phoneme processor (G2P) based on a dictionary of pronunciations. The \textit{X-Sampa} phoneme set \cite{wells1995} has been used along all the experiments. 80-mel spectrograms have been extracted from the recordings (using 50ms window length, 12.5ms skips and 1024-point FFTs), and then they have been standardized. We held out a set composed of 5000 utterances, for development purposes. Along the following experiments, we describe various test sets. These test sets are always disjoint from the training set described above. We have trained that model for 2.5M training steps using \textit{Adam} optimizer \cite{kingma14} with default hyper-parameters,  a learning rate of $5\cdot 10^{-4}$, a \textit{minibatch} size of 128 and using a machine with 8 \textit{Nvidia V100 16GB Tensor Core GPUs}. 

Both the phonetic and the acoustic encoder transformer backbones are built with 3 layers of 8 attention heads each and a dropout rate of 0.1. The encoders operate with hidden and output spaces of size $\mathbb{R}^{256}$. The \textit{LSTM} integrator head, which is shared between both encoders, is a single layer \textit{LSTM} with 1024 units, which corresponds to the size of the output embeddings (similar to \textit{CLIP}). The last output state of the \textit{LSTM} is used as the final \textit{SCRAPS} vector. We used a temperature of $T=1.0$ during all the training and concluded that our training process is stable with that setting.

Contrastive models are difficult to evaluate in an objective manner \cite{robinson2021}. In our case, just taking an unseen set of utterances, computing the \textit{SCRAPS} vectors from a random sample of the spectrograms and  phonetic sequences and measuring how well the model discriminates between matching and non-matching pairs is not informative, due to the fact that the examples in the \textit{minibatches} are very dissimilar among them (i.e. when the  samples are chosen at random to build the \textit{minibatch} it may be too easy to discriminate between positive and negative pairs). We have conducted this exercise with a held-out test set of around 6000 samples, and achieved an Equal-Error-Rate (EER) in the order of $10^{-4}$ and Area Under the Receiver Operating Curve (\textit{AUC-ROC}) in the order of $1 - 10^{-7}$. Handcrafting a test set by enrolling challenging negative samples is a possible alternative \cite{cao2022}, but it relies on a criterion of difficulty that is not easy to define and that is prone to subjective biases. This difficulty motivates the exploration of more creative methods for understanding the performance of our model. 

We propose a sensitivity and robustness analysis to study how does the model react against perturbation of the input data (referred hereafter as corruption). In this analysis, we corrupt either the spectrogram or the phonetic sequence and measure how the results vary due to that perturbation. 
%The methods used to corrupt the spectrograms and the phonetic sequences are described below
We use the following methods to corrupt spectrograms or phonetic sequences:

\begin{itemize}
	\item \textbf{Random phonemes substitution}: substitute a portion of phonemes in the phonetic sequence at random.
	\item \textbf{Spectrogram \textit{Gaussian} noise}: add different amounts of \textit{Gaussian} noise to spectrogram. The noise is added according to Equation \ref{eq:noise_addition}, where $\mathbf{M}$ is the spectrogram, $\mathbf{N}$ is the noise and $\alpha \in [0,1]$ is the varying parameter that controls the amount of noise.
	\item \textbf{Spectrogram mix}:  analogous to the \textit{Gaussian} noise method, but replacing the noise with another random sampled spectrogram from the \textit{minibatch}. Equation \ref{eq:noise_addition} is used to mix the original and the noise sources.
\end{itemize}

\begin{equation} \label{eq:noise_addition}
	\mathbf{\hat M} = (1-\alpha) \cdot \mathbf{M} + \alpha \cdot \mathbf{N}.
\end{equation}

\begin{table*}[h] \footnotesize \setlength{\tabcolsep}{3pt}
\vspace{10pt} 
\caption{Robustness (\textit{AUC-ROC}) and sensitivity (\textit{drop} and \textit{lift}) results of \textit{SCRAPS} at different levels of corruption. The second row of the table indicates the method used to corrupt the inputs in each case. All the results are represented as the mean $\pm$ the 95\% confidence interval. }	\label{tab:results}
\centering
\begin{tabular}{r|l|ll|l|ll|l|ll}
	\toprule
	 \multicolumn{1}{r|}{Space} &                                   \multicolumn{6}{c|}{Acoustic}                                   &          \multicolumn{3}{c}{Phonetic}          \\ \midrule
	\multicolumn{1}{r|}{Method} &      \multicolumn{3}{c|}{Mix spectrograms}      &       \multicolumn{3}{c|}{\textit{Gaussian} noise}       &    \multicolumn{3}{c}{Random substitution}     \\\midrule
	\multicolumn{1}{r|}{Metric} & \textit{AUC-ROC}           & \textit{Drop} (\%)      & \textit{Lift} (\%)      & \textit{AUC-ROC}           & \textit{Drop} (\%)      & \textit{Lift} (\%)      & \textit{AUC-ROC}           & \textit{Drop} (\%)      & \textit{Lift} (\%)     \\
	            Corruption (\%) &               &                &                &               &                &                &               &                &  \\ \midrule
	                        5.0 & 1.00$\pm$0.00 & 46.01$\pm$2.88 & 53.99$\pm$2.88 & 1.00$\pm$0.00 & 43.49$\pm$2.86 & 56.51$\pm$2.86 & 1.00$\pm$0.00 & 52.78$\pm$2.88 & 8.94$\pm$1.65 \\
	                       10.0 & 1.00$\pm$0.00 & 49.31$\pm$2.89 & 50.69$\pm$2.89 & 1.00$\pm$0.00 & 45.14$\pm$2.87 & 54.86$\pm$2.87 & 1.00$\pm$0.00 & 76.91$\pm$2.43 & 7.20$\pm$1.49 \\
	                       20.0 & 1.00$\pm$0.00 & 53.21$\pm$2.88 & 46.79$\pm$2.88 & 1.00$\pm$0.00 & 47.83$\pm$2.88 & 52.17$\pm$2.88 & 1.00$\pm$0.00 & 91.06$\pm$1.65 & 2.60$\pm$0.92 \\
	                       40.0 & 1.00$\pm$0.00 & 79.43$\pm$2.33 & 20.57$\pm$2.33 & 1.00$\pm$0.00 & 63.54$\pm$2.78 & 36.46$\pm$2.78 & 0.97$\pm$0.01 & 96.44$\pm$1.07 & 1.22$\pm$0.63 \\
	                       60.0 & 0.76$\pm$0.02 & 95.23$\pm$1.23 & 4.77$\pm$1.23  & 1.00$\pm$0.00 & 92.10$\pm$1.56 & 7.90$\pm$1.56  & 0.85$\pm$0.02 & 98.44$\pm$0.72 & 0.87$\pm$0.54 \\
	                       80.0 & 0.54$\pm$0.03 & 99.57$\pm$0.38 & 0.43$\pm$0.38  & 0.80$\pm$0.02 & 99.22$\pm$0.51 & 0.78$\pm$0.51  & 0.71$\pm$0.03 & 99.48$\pm$0.42 & 0.35$\pm$0.34 \\
	                       90.0 & 0.52$\pm$0.03 & 99.65$\pm$0.34 & 0.35$\pm$0.34  & 0.65$\pm$0.03 & 99.13$\pm$0.54 & 0.87$\pm$0.54  & 0.62$\pm$0.03 & 99.57$\pm$0.38 & 0.35$\pm$0.34 \\
	                       95.0 & 0.51$\pm$0.03 & 99.65$\pm$0.34 & 0.35$\pm$0.34  & 0.57$\pm$0.03 & 99.39$\pm$0.45 & 0.61$\pm$0.45  & 0.59$\pm$0.03 & 99.83$\pm$0.24 & 0.17$\pm$0.24 \\ \bottomrule
\end{tabular}
\end{table*}
\subsection{Sensitivity} \label{sec:sensitivity}

We take a random sample of 1152 matching phonetic sequences and spectrograms from the test set, and calculate their \textit{SCRAPS} score. We apply the corruption methods described above, one by one, and calculate their corresponding \textit{SCRAPS} scores.  Finally, for each of the tested corruption proportions we calculate the percentage of cases where the \textit{SCRAPS}  score of the corrupted examples \textit{drops} or \textit{lifts} by any amount, with respect to the original score (without corruption). High sensitivity to phonetic corruption is desirable to spot if utterances and transcriptions match. Hence, a good model would have an early raise of \% of drops, and none/very few lifts. However, for acoustic corruption, high sensitivity is bad. Corrupting the acoustics to change the content is not an easy task. We instead noised the utterances with Gaussian noise, or by mixing in another utterance. With those methods a good model would have low sensitivity to acoustic corruption (i.e., ability to match phonetic and acoustic content in adverse conditions). 

\begin{figure}[h]
	\centering
	\includegraphics[width=1.0\linewidth]{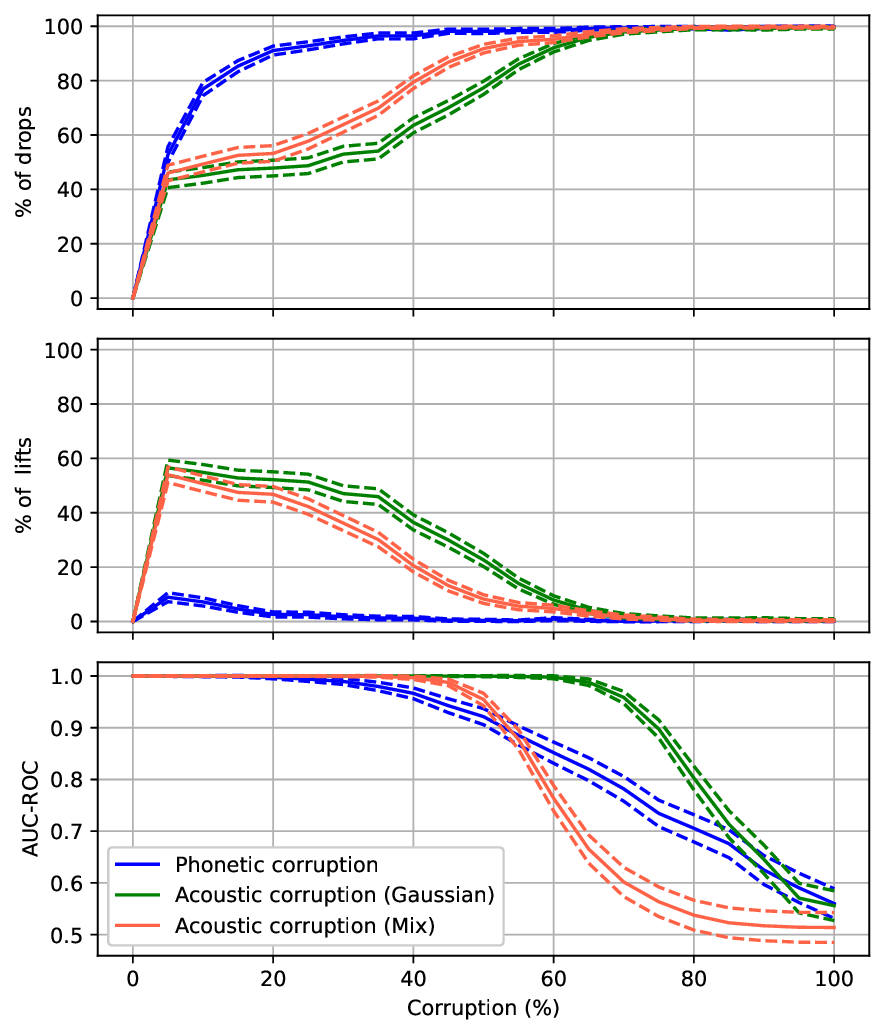}
	\caption{Sensitivity and robustness results. Top and middle charts represent the percentage of \textit{drops} (top) and \textit{lifts} (middle) in \textit{SCRAPS} score as a function of different levels of corruption in the input phonemes score. Bottom chart shows the robustness evaluation results using \textit{Gaussian} noise and spectrogram mix methods. 
% 	The dashed lines represent the 95\% interval over the mean. 
	Dashed lines represent the 95\% confidence intervals.}
	\label{fig:results}
\end{figure}

The results are represented visually in Figure \ref{fig:results}-top and \ref{fig:results}-middle. Table \ref{tab:results} shows the percentage of \textit{drops} and \textit{lifts} in \textit{SCRAPS} scores after corruption for a set of specific corruption amounts.

In Table \ref{tab:results} we observe that in the $91.06\%$ of the examples where at least $20\%$ of the phonemes are corrupted, the \textit{SCRAPS} model reacts by decreasing the score of the match, while in $2.62\%$ of the cases the model score increases.

\begin{figure}[h]
	\centering
	\includegraphics[width=1.0\linewidth]{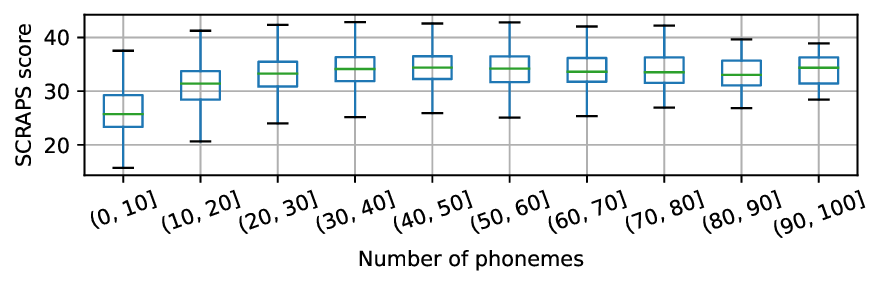}
	\caption{Distribution of \textit{SCRAPS} scores with respect to the lengths of the phonetic sequences.}
	\label{fig:lengths}
\end{figure}

Finally, we also study the effect that the number of phonemes in the sequences has in the \textit{SCRAPS} score (see figure \ref{fig:lengths}), observing that the distribution of \textit{SCRAPS} score remains constant when the number of phonemes is greater than 20. 

\subsection{Robustness} \label{sec:robustness}

Another important feature to test on the \textit{SCRAPS} model is how robust it is against perturbations. In the worst case scenario, this model would only work on conditions that are very similar to those of the training data, and completely fail if the data characteristics change slightly, which is undesirable. To measure the robustness, we have calculated the \textit{AUC-ROC} score over the \textit{SCRAPS} probabilities of each \textit{minibatch}, considering the matching pairs as positive examples and the non-matching pairs as negative examples. We have calculated that score after corrupting the inputs in different quantities.  These measurements have been performed over 1152 random sampled examples from the test set.

The results of this experiment for the two types of noise are represented in Figure \ref{fig:results}-bottom where the amount of noise has been represented in the $x$ axis and the \textit{AUC-ROC} in the $y$ axis. Table \ref{tab:results} provides the \textit{AUC-ROC} scores achieved for a set of manually selected noise amounts ($\alpha$ in Equation \ref{eq:noise_addition}). As we can observe, the \textit{SCRAPS} model is robust against different levels of noise depending on its nature. For \textit{Gaussian} noise, the performance degrades a $10\%$ or more when adding noise with $\alpha\geq 0.75$, However, for spectrogram mixing, the performance drops at least a $10\%$ when $\alpha\geq 0.52$. 

An interesting observation arises when looking at Figure \ref{fig:results}-bottom: the \textit{Gaussian} noise curve (green) never reaches the random performance level (even for $\alpha=1.0$, point in which the spectrograms have no information at all, the \textit{AUC-ROC} is significantly greater than 0.5). We hypothesize that this is due to the fact that despite not providing explicit information to the acoustic transformer, it extracts information from the positional encoding to later determine if the length of the acoustic input is feasible for the phonetic sequence provided.

\subsection{Covariate shift}
Looking at the results of subsection \ref{sec:sensitivity}, one could think that the observed effects may be due to covariate shift induced by the corruption methods described.  Four additional experiments have been conducted. These experiments involved more sophisticated corruption techniques such as word swaps, cut-offs, accent switching, and denoising. In all cases, the \textit{SCRAPS} scores were affected by the corruptions, hence the likelihood of covariate shift being the cause of the initial drop in scores was determined to be minimal. Appendix  
A,%\ref{sec:cshift}
 which we included in the preprint version of this manuscript\footnote{https://arxiv.org/abs/2307.12445} due to length constraints, describes the findings of these experiments more in detail.

\subsection{Open source evaluation baselines}
With the aim of facilitating future benchmarking and reproducibility, we have included two extra evaluations using the \textit{Common Voice} test set. In the first experiment we measure the correlation between \textit{SCRAPS} scores and \textit{Whisper} base perplexity using the invalidated subset of the \textit{Common Voice} dataset, observing a Pearson correlation of -0.60. This indicates a significant level of agreement between \textit{SCRAPS} and \textit{Whisper}, which establishes a first reproducible benchmark.

The second experiment replicates the sensitivity analysis against natural noise (denoising) described previously (Appendix 
A). %\ref{sec:cshift}
 This time, the open source \textit{Common Voice} test set is used, which contains approximately 16K utterances. After denoising, the \textit{SCRAPS} scores increase by 2.94 points, with 72.79\% of the utterances showing a score increase, confirming that \textit{SCRAPS} scores significantly increase after denoising the utterances. 

Due to length restrictions, these results have been included in Appendix 
B %\ref{sec:osource} 
of the preprint\footnotemark[1].

\subsection{Examples}

Aside from quantitative results, qualitative ones are often informative. This section provides a set of cherry-picked examples where we use \textit{SCRAPS} to retrieve the spectrograms with highest score in our test set given an input transcription. For that, we score a random sample of the test set containing 7124 pairs (spectrogram, phonetic sequence) and calculate the pairwise dot product between acoustic and phonetic vectors. The resulting matrix has shape $7124 \times 7124$. Then we normalize that matrix and do a lookup to find the closest spectrograms given a sentence. In the paragraphs below, the first sentence highlighted in bold represents the query sentence, and the ones below are the transcriptions corresponding with the top 3 spectrograms with highest probability. The numerical probability value, rounded to the 2nd decimal place, is indicated between parentheses before the sentences. In the cases where one of the top 3 closest spectrograms corresponds to the actual matching spectrogram for the reference sentence, we have highlighted its transcription with a star (*).\\

% \mbox{}  % empty space for first column
%\newpage % start next column

{\footnotesize 
\textbf{Astro dance.}\\
$*$(0.97) Astro dance.\\
(0.03) Astro, dance. \\
(0.00) Astro dance for me. 

\textbf{Alexa, end this call on my phone.}\\
$*$(1.00) Alexa, end this call on my phone.\\
(0.00) Alexa, unpair my fire phone. \\
(0.00) Alexa, end this meeting for all. 

\textbf{Echo, keep reading.}\\
$*$(0.74) Echo, keep reading.\\
(0.25) Hey Alexa keep reading. \\
(0.01) Echo, keep going.

\textbf{Alexa, disconnect my fire phone.}\\
$*$(0.88) Alexa, disconnect my fire phone.\\
(0.04) Alexa disconnect the phone. \\
(0.03) Alexa disconnect my cell phone. 

\textbf{Alexa, do i have any appointments on June nineteenth.}\\
$*$(0.98) Alexa, do i have any appointments on June nineteenth. \\
(0.01) Alexa do i have any appointments on February twenty fifth. \\
(0.00) Alexa, do i have any appointments at seven fifty five AM. \\
}

One of the most important observations by looking at the examples above is that in some cases the commas in the transcriptions (corresponding to pauses in the associated spectrograms, given that the commas were annotated through forced alignment \cite{mcauliffe2017}) have a notable effect in the \textit{SCRAPS} probability. We have analysed that effect in more detail and observed that when removing the commas from a pair that originally had a comma, the \textit{SCRAPS} score \textit{drops} a $5.49\% \pm 0.38\%$ with respect to the original value (as per the 95\% confidence interval over the mean, calculated with 768 samples). Through an informal listening test, we have confirmed that the commas usually correspond to long pauses in the spectrogram.

\begin{figure}[h!]
	\centering
	\includegraphics[width=1.0\linewidth]{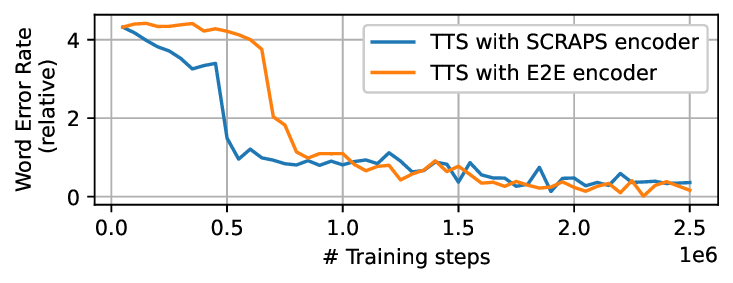}
	\caption{Word Error Rate (WER) of the TTS models, relative to the WER over the test set recordings. The two lines represent how the intelligibility of the model improves as the model trains for the architectures with pretrained SCRAPS and with E2E phonetic encoder. The WER has been measured with a pretrained ASR model over a set of 512 sentences unseen during training, synthesized every 25,000 training steps. The WER represented in the figure is relative to the recordings WER.
	}
	\label{fig:werr_emb}
\end{figure}

\subsection{Downstream application 1: pretrained phonetic embeddings for speech generation} \label{sec:emb}

For this application, we have trained an autoregressive multispeaker text-to-speech model with attention, similar to Transformer TTS.  We used a proprietary dataset with 180,000 hours of de-identified en-US recordings. Around 2/3 of the dataset is automatically annotated by an ASR model, the rest of utterances have been manually annotated. The evaluation set, used to generate the figure 4, is composed of 10,000 sentences, randomly held out from the training set. 

The baseline architecture consists of three modules: a phonetic encoder, an acoustic decoder and a speaker embedding. The acoustic decoder is autoregressive and attends to the output of the phonetic encoder and the speaker embedding. Then we have substituted the phonetic encoder by a \textit{SCRAPS} pretrained phonetic encoder (only the transformer backbone, not the \textit{LSTM} integrator). The test metrics evolution during training are represented in Figure \ref{fig:werr_emb}. In the figure, we observe that both architectures get a very similar final performance, but when using \textit{SCRAPS}, the model converges much faster (the initial error drop happens 250,000 steps earlier in the case of the model that uses the pretrained \textit{SCRAPS}). Apart from the clear faster convergence, this observation allows us concluding that the \textit{SCRAPS} transformer latent space is rich enough to be used as a pretrained module to speed up Text-To-Speech (TTS) model training.

\subsection{Downstream application 2: text-less intelligibility evaluation for voice conversion systems} \label{sec:vc}

Intelligibility performance is a key metric for Voice Conversion (VC) systems \cite{vcc_2018}. Usually, WER is used to assess such performance. However, it is relatively costly: apart from ground-truth texts, one needs to transcribe VC audio using humans or ASR. In case of using an ASR system, it needs to be trained on the target language with sufficient amount of data in order to be robust enough to bypass the need for human annotations. For these reasons, there is a need for cheaper alternatives that, if possible, do not require any text. 

In this application we study the use of \textit{SCRAPS} as a text-less intelligibility metric. For that, we used a VC model trained over a propietary dataset with 60k hours of de-identified en-US recordings, manually annotated by humans. Each of these recordings are paired with a synthetically generated audio of a target speaker, to train the many-to-one voice conversion model. These syntheses have been generated using a similar technique of the one described in this paper \cite{gabrys2022}. As mentioned in the section 4.6 of the manuscript, the evaluation set is composed of 3500 utterances, which were held out from the training set. Although \textit{SCRAPS} is trained to match a sequence of phonemes to the corresponding audio, at inference time it can also be used to compute correspondence between two audio files without requiring any text. In this scenario, the \textit{SCRAPS} score is computed between vectors of synthetic audio (VC) and source audio (pre-conversion). We refer to this metric as \textit{SCRAPS(VC, source)}.

We evaluate usability of \textit{SCRAPS(VC, source)} as an intelligibility metric by testing how it relates to WER computed over human annotations (h-WER). We run an en-US any-to-one VC system (similar to \cite{parrotron}) over a test set of 3532 recordings. Then, we compute h-WER by gathering human transcriptions on VC audio and comparing it to ground-truths. Finally, \textit{SCRAPS(VC, source)} is computed using only VC and source (pre-conversion) audios. We compute the \textit{Spearman} correlation between \textit{SCRAPS(VC, audio)} and h-WER (Table \ref{tab:wer_corr}). For comparison, two alternative metrics are computed: (1) ASR-WER, where we use a proprietary ASR system to replace human transcriptions, (2) \textit{SCRAPS(VC, text)} that uses ground-truth text instead of input audio. The proposed metric obtains a negative correlation of $-0.1938$, which is significantly lower (in absolute values) than \textit{SCRAPS(VC, text)} ($-0.3232$) and ASR ($0.6197$). Since h-WER has a bounded distribution at 0, with a large proportion of samples ($69\%$) having exactly h-WER=0, while \textit{SCRAPS} scores are unbounded, we also compute correlation using only samples with h-WER$>$0 (third column).
%However, when we look only at utterances with h-WER$>$0 (third column)
As can be observed, the differences in correlation between metrics are then smaller, with no perceivable difference between \textit{SCRAPS(VC, source)} and \textit{SCRAPS(VC, text)} scores. This suggests that \textit{SCRAPS(VC, source)} more often incorrectly assigns low score to high-intelligibility (h-WER=0) samples than \textit{SCRAPS(VC, text)}, which could originate from \textit{SCRAPS} not being trained with pairs of audio and being less robust to slight differences between them, even if content in both is the same.
This experiment shows there is a statistically significant correlation between SCRAPS scores and h-WER, although using a robust ASR is still the strongest automatic solution, despite the higher computational cost.

\begin{figure}[h!]
	\centering
	\includegraphics[width=1.0\linewidth]{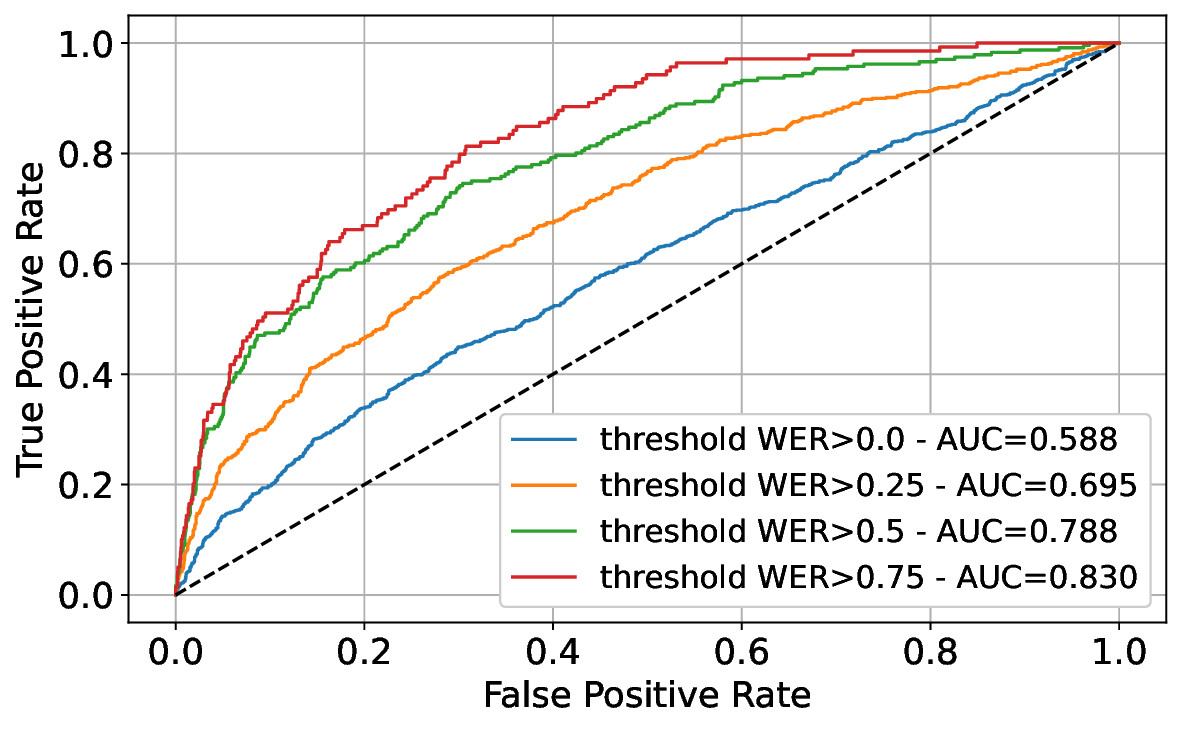}
	\caption{ROC of h-WER$>$T classification using SCRAPS(VC, source) scores for different thresholds T.}
	\label{fig:roc_scraps}
\end{figure}

We also test if \textit{SCRAPS(VC, source)} can be used to find high-WER VC samples, which could serve as a post-filter to remove low-intelligibility VC outputs. The results can be seen in Figure \ref{fig:roc_scraps}, where four h-WER thresholds were used to binarize samples into "low" and "high" h-WER samples. While \textit{SCRAPS(VC, source)} is not robust enough to classify h-WER$>$0 samples without supervision (AUC=$0.580$), it might be sufficient enough to act as a filter for removing the worst cases, \textit{i.e.} h-WER$>$0.75 (AUC=$0.830$), thus reducing the computational cost of the filtering process.

\begin{table}[h!] %\footnotesize %\setlength{\tabcolsep}{2.3pt}
\vspace{10pt} 
\centering
\small
\caption{Spearman correlation between automatic intelligibility metrics and human WER (h-WER). Results on all utterances (second column) and utterances with h-WER$>$0 (third column) are shown.}
\label{tab:wer_corr}
\begin{tabular}{l|ll}
\toprule
\textbf{Metric}   & \textbf{\begin{tabular}[c]{@{}l@{}}Correlation \\ (all)\end{tabular}} & \textbf{\begin{tabular}[c]{@{}l@{}}Correlation\\ (h-WER$>$0)\end{tabular}} \\ \midrule
ASR-WER           & 0.6197                                                             & 0.7064                                                                                 \\
\textit{SCRAPS(VC, text)}  & -0.3232                                                            & -0.4812                                                                                \\ 
\textit{SCRAPS(VC, source)} & -0.1938                                                            & -0.4740                                                                                \\ \bottomrule
\end{tabular}
\end{table}

\subsection{Ablation study}

% BASE MODEL
\begin{table*}[h!] \footnotesize \setlength{\tabcolsep}{2.3pt}
	\vspace{10pt} 
	\caption{Results of the ablation study, showing the robustness (\textit{AUC-ROC}) and sensitivity (\textit{drop} and \textit{lift}) numbers of the ablated versions of \textit{SCRAPS} at different levels of corruption. All the results are represented as the mean $\pm$ the 95\% confidence interval. The cases where the results differ significantly from the originally defined version of \textit{SCRAPS} have been highlighted in bold. They have been colored in blue in case of representing a negative deviation over the baseline (ablated $<$ baseline) or in red if the deviation is positive (ablated $>$ baseline).}	\label{tab:ablation}
	\centering
	\begin{tabular}{r|r|c|cc|c|cc|c|cc}
		\toprule
		         &  \multicolumn{1}{r|}{Space} &                                                                                                     \multicolumn{6}{c|}{Acoustic}                                                                                                      &                         \multicolumn{3}{c}{Phonetic}                         \\ \midrule
		         & \multicolumn{1}{r|}{Method} &                                \multicolumn{3}{c|}{Mix spectrograms}                                 &                                               \multicolumn{3}{c|}{\textit{Gaussian} noise}                                               &                   \multicolumn{3}{c}{Random substitution}                    \\ \midrule
		         & \multicolumn{1}{r|}{Metric} &     \textit{AUC-ROC}     &                \textit{Drop} (\%)                 &                \textit{Lift} (\%)                &                 \textit{AUC-ROC}                  &                \textit{Drop} (\%)                 &                 \textit{Lift} (\%)                 &                 \textit{AUC-ROC}                 &    \textit{Drop} (\%)     &    \textit{Lift} (\%)    \\
		Ablation &                     Corruption (\%) &                 &                                          &                                         &                                          &                                          &                                           &                                         &                  &                 \\
		         &                  &                 &                                          &                                         &                                          &                                          &                                           &                                         &                  &                 \\ \midrule
		  Parameters &                         5.0 &  1.00$\pm$0.00  &              49.57$\pm$2.89              &            {50.43$\pm$2.89}             &              1.00$\pm$0.00               &              45.57$\pm$2.88              &             {54.43$\pm$2.88}              &              1.00$\pm$0.00              &  52.00$\pm$2.89  &  9.72$\pm$1.71  \\
		 sharing &                        10.0 &  1.00$\pm$0.00  &              49.74$\pm$2.89              &            {50.26$\pm$2.89}             &              1.00$\pm$0.00               &              45.31$\pm$2.87              &              54.69$\pm$2.87               &              1.00$\pm$0.00              &  75.95$\pm$2.47  &  8.16$\pm$1.58  \\
		         &                        20.0 &  1.00$\pm$0.00  &              56.77$\pm$2.86              &            {43.23$\pm$2.86}             &              1.00$\pm$0.00               &              47.57$\pm$2.88              &              52.43$\pm$2.88               &              1.00$\pm$0.00              &  90.36$\pm$1.70  &  3.30$\pm$1.03  \\
		         &                        40.0 &  1.00$\pm$0.00  &              81.51$\pm$2.24              &            {18.49$\pm$2.24}             &              1.00$\pm$0.00               &              61.20$\pm$2.81              &              38.80$\pm$2.81               &              0.96$\pm$0.01              &  96.18$\pm$1.11  &  1.48$\pm$0.70  \\
		         &                        60.0 & {0.77$\pm$0.02} &              96.53$\pm$1.06              &             {3.47$\pm$1.06}             &              1.00$\pm$0.00               &              90.89$\pm$1.66              &               9.11$\pm$1.66               &              0.85$\pm$0.02              &  98.52$\pm$0.70  &  0.78$\pm$0.51  \\
		         &                        80.0 &  0.54$\pm$0.03  &              99.57$\pm$0.38              &              0.43$\pm$0.38              &              0.79$\pm$0.02               &              98.96$\pm$0.59              &               1.04$\pm$0.59               &              0.71$\pm$0.03              &  99.39$\pm$0.45  &  0.43$\pm$0.38  \\
		         &                        90.0 &  0.52$\pm$0.03  &              99.74$\pm$0.29              &             {0.26$\pm$0.29}             &              0.62$\pm$0.03               &             {99.39$\pm$0.45}             &              {0.61$\pm$0.45}              &             {0.63$\pm$0.03}             &  99.65$\pm$0.34  & {0.26$\pm$0.29} \\
		         &                        95.0 & {0.52$\pm$0.03} &              99.65$\pm$0.34              &              0.35$\pm$0.34              &              0.57$\pm$0.03               &              99.31$\pm$0.48              &               0.69$\pm$0.48               &             {0.60$\pm$0.03}             &  99.83$\pm$0.24  &  0.17$\pm$0.24  \\ \midrule
		    \textit{LSTM} &                         5.0 &  1.00$\pm$0.00  &              48.00$\pm$2.89              &            {52.00$\pm$2.89}             &              1.00$\pm$0.00               &              51.30$\pm$2.89              & \textcolor{blue}{\textbf{48.70$\pm$2.89}} &              1.00$\pm$0.00              &  51.56$\pm$2.89  & 10.16$\pm$1.74  \\
		         &                        10.0 &  1.00$\pm$0.00  &              49.74$\pm$2.89              &            {50.26$\pm$2.89}             &              1.00$\pm$0.00               &              49.74$\pm$2.89              &             {50.26$\pm$2.89}              &              1.00$\pm$0.00              &  75.26$\pm$2.49  &  8.85$\pm$1.64  \\
		         &                        20.0 &  1.00$\pm$0.00  &              54.77$\pm$2.87              &            {45.23$\pm$2.87}             &              1.00$\pm$0.00               & \textcolor{red}{\textbf{59.29$\pm$2.84}} & \textcolor{blue}{\textbf{40.71$\pm$2.84}} &              1.00$\pm$0.00              &  90.62$\pm$1.68  &  3.04$\pm$0.99  \\
		         &                        40.0 &  1.00$\pm$0.00  &              80.03$\pm$2.31              &            {19.97$\pm$2.31}             &              1.00$\pm$0.00               & \textcolor{red}{\textbf{92.97$\pm$1.48}} & \textcolor{blue}{\textbf{7.03$\pm$1.48}}  &             {0.98$\pm$0.01}             &  96.44$\pm$1.07  &  1.22$\pm$0.63  \\
		         &                        60.0 &  0.76$\pm$0.02  & \textcolor{red}{\textbf{98.09$\pm$0.79}} & \textcolor{red}{\textbf{1.91$\pm$0.79}} & \textcolor{blue}{\textbf{0.77$\pm$0.02}} & \textcolor{red}{\textbf{99.22$\pm$0.51}} & \textcolor{blue}{\textbf{0.78$\pm$0.51}}  & \textcolor{red}{\textbf{0.90$\pm$0.02}} &  98.52$\pm$0.70  & {0.78$\pm$0.51} \\
		         &                        80.0 &  0.54$\pm$0.03  &              99.74$\pm$0.29              &             {0.26$\pm$0.29}             & \textcolor{blue}{\textbf{0.54$\pm$0.03}} &              99.57$\pm$0.38              &              {0.43$\pm$0.38}              &              0.76$\pm$0.02              & {99.57$\pm$0.38} & {0.26$\pm$0.29} \\
		         &                        90.0 & {0.53$\pm$0.03} &              99.74$\pm$0.29              &             {0.26$\pm$0.29}             & \textcolor{blue}{\textbf{0.53$\pm$0.03}} &              99.48$\pm$0.42              &              {0.52$\pm$0.42}              &              0.68$\pm$0.03              &  99.57$\pm$0.38  & {0.35$\pm$0.34} \\
		         &                        95.0 & {0.53$\pm$0.03} &              99.91$\pm$0.17              &             {0.09$\pm$0.17}             & \textcolor{blue}{\textbf{0.52$\pm$0.03}} &              99.57$\pm$0.38              &              {0.43$\pm$0.38}              &              0.64$\pm$0.03              &  99.65$\pm$0.34  &  0.35$\pm$0.34  \\ \bottomrule
	\end{tabular}
\end{table*}

In this subsection, we question some of the design decisions of the \textit{SCRAPS} architecture and test the impact of ablating them in terms of sensitivity and robustness. The aspects included in this ablation study are the following ones:

\begin{itemize} 
	\item Head parameters sharing: this feature was included to try to bias the acoustic and phonetic transformer towards producing compatible representations. However, there is a risk that by doing that, the representation power of the encoders decreases. 
	\item \textit{LSTM} integrator modules: they may be superfluous. In such case, the output of the acoustic and phonetic encoders is the vector corresponding to the sequence length at the output of the transformer. 
\end{itemize}

To do these experiments we trained two new models with the described ablated components, for the same number of training steps as the full model described before. The results, included in Table \ref{tab:ablation}, show that in the majority of metrics there is no significant difference between the baseline and the ablated versions. More specifically, from the table we can conclude that sharing parameters of the \textit{LSTM} integrators is safe, as it does not have a significant effect in the performance, compared with the baseline. However, the \textit{SCRAPS} version without integrators seems to be significantly less robust to \textit{Gaussian} noise than the rest, showing large differences in \textit{AUC-ROC} score. This may imply that the transformer is not large enough to handle noisy inputs and that the \textit{LSTM} module at the output of the transformer helps making the output more robust. 

\vspace{-5pt}
\section{Discussion}
The \textit{SCRAPS} model has shown the ability to model a rich common latent space between phonetic and acoustic spaces, which is interesting for speech generation and recognition tasks. This task is not trivial, as it involves implicitly learning the underlying allophonic and phonetic joint probability distribution, which is the root difficulty of speech technologies, while learning to ignore non-related information such as prosody, background, non-verbal sounds and other stationary aspects such as speaker id or different accents. 

We have proved that the \textit{SCRAPS} model has high sensitivity to small changes in phonetic sequences, and at the same time it is robust against high levels of noise, independently of its nature (stationary and non-stationary). One of the most surprising findings has been \textit{SCRAPS} robustness in the face of concurrent speech, showing an AUC-ROC of $1.00$ up to 60\%-40\% overlap of primary and secondary spectrograms, respectively (see Figure \ref{fig:results}-bottom). 

The \textit{SCRAPS} model can be seen as an implicit speech generation-recognition system. Given a mel-spectrogram, one can search for the sentence that better matches that spectrogram, transforming the system in an automatic speech recognizer. On the other side, given a sentence, one could look for the spectrogram that better matches the input sentence, thus converting the system into a speech generator. Obviously the search processes makes these approaches very computationally expensive, but this interpretation of \textit{SCRAPS} helps devise further applications.

Finally, on the practical side, \textit{SCRAPS} brings its value when used as a tool for a downstream application. There are many potential use cases for this model, two of which we explored in sections \ref{sec:emb} and \ref{sec:vc}. Further research lines could explore other applications. Below, we state some of the most interesting examples, but there may be many more.

\begin{itemize}
    \item Transcription quality assurance: machine learning models are sensitive to the quality of the training data. In speech generation and recognition applications, the datasets often contain transcription problems at different levels: from mistaken words to poor annotation (e.g. not containing hesitations or background events). \textit{SCRAPS} can be trained with high quality data and used to score other datasets to spot transcription issues. 
    \item Grapheme to phoneme mapping enhancement: speech datasets are usually composed of pairs of audio clips and sentences. The sentences are mapped to the phonetic sequences using grapheme to phoneme (G2P) systems, which often rely on a word-level dictionary of pronunciations \cite{taylor2009}. Although this approach usually works in practice, it is prone to errors when a speaker in the data set uses non-normative pronunciations, or in the presence of homographs \cite{ploujnikov2022}. \textit{SCRAPS} could potentially be used to find the most probable phoneme, that better matches the allophones present in the associated recording, allowing to have a higher quality automatic annotation and hence a potentially better model.
    \item Intelligibility optimization: intelligibility is an important metric for speech synthesis systems, although it is not the only one (other examples are signal quality, speaker id match or naturalness). Usually, these models do not directly optimize these objectives, but others such as the maximum likelihood \cite{bishop2006}, that indirectly optimize the previously mentioned ones. It has been repeatedly proven that the those indirect objectives do not always correlate with the perceptual ones \cite{goodfellow2016}. \textit{SCRAPS} model, as a consequence of being a fully differentiable system, offers the possibility of being used as a loss function to directly optimize the match between the target transcription and the content of the spectrogram, or in other words, to directly maximize intelligibility. This could be achieved by plugging the \textit{SCRAPS} encoders (frozen) in a speech generation architecture and maximizing the dot product between the target phonetic sequence and the synthesized spectrogram. This could be done in combination with the traditional objectives. Alternatively, one could also use \textit{SCRAPS} to monitor the evolution of intelligibility during training in order to, for instance, implement early stopping mechanisms.
\end{itemize}
\vspace{-15pt}

\section{Conclusions}
\vspace{-5pt}
We showed how by using a \textit{CLIP}-like architecture we are able to model a latent space that connects phonetic and acoustic domains. The experiments provide evidence that the embeddings are robust to different kinds of noise, with no degradation up to 60\% of corruption when using \textit{Gaussian} noise and 40\% when dealing with concurrent speech. At the same time, we observed that \textit{SCRAPS} reacts decreasing the score, in $91.06\%$ of the examples where at least $20\%$ of the phonemes are corrupted. Finally, we provided two examples of downstream applications that show that \textit{SCRAPS} is a useful component in several downstream applications, and discussed future research lines.

\bibliography{references.bib}

\end{document}